\documentclass[prd,twocolumn,tightenlines]{revtex4}
\usepackage{amsmath}
\usepackage{amssymb}
\usepackage{graphicx}
\usepackage{MnSymbol}

\begin{document}

\title{ Quantum Corrections to Multi-Quanta Higgs-Bags in the Standard Model }
\author{Marcos P. Crichigno and Edward Shuryak}
\affiliation{Department of Physics and Astronomy \\
Stony Brook University, Stony Brook NY 11794 USA }
\date{\today}

\begin{abstract}
We argue that the Standard Model contains stable bound states with a sufficiently
large number $N$ of heavy quanta -- top quarks $t, \bar t$ and gauge bosons $W,Z
$ -- of the form of collective ``bags", with a strongly depleted value
of the Higgs VEV inside. More specifically, we study one-loop quantum
corrections to a generic model of them, assuming ``quanta" are described by
a complex scalar field. We follow the practical formalism developed  by Farhi \textit{et al.} for the $N=1$
case, i.e. one particle in a bag, who found that for a very large  Yukawa coupling the classical bags
are destabilized by quantum effects. We instead study the problem with a coupling
constant in the range of the Standard Model for a large number of quanta $N=50..5000$. We calculated both
classical and one-loop effects and  found that for such bags quantum
corrections are small, of the order of a few percents or less. \newline
\end{abstract}

\maketitle

\address {Department of Physics and Astronomy, State University of New York,
Stony Brook, NY 11794}

\section{Introduction}

   Coherent bound states of a large number of  bosonic quanta are known as ``semiclassical solitons".  An important subclass of them -- \emph{topological solitons} -- play a significant role in physics in general and in the dynamics of gauge
theories in particular.  Electroweak sphalerons, for instance, were
extensively studied in the context of baryon number violation in the Standard
Model (SM). Instantons drive the breaking of chiral $U(1)$
and $SU(N_{f})$ symmetries in QCD, while dyons/monopoles are believed to be
related to the phenomenon of color confinement.

 The objects we will study in this
paper belong to the less famous subclass, namely \emph{%
``non-topological} solitons", which have remained somehow in the
shadow of their topological relatives, although also studied --
 as pure theoretical constructions --  for a long time. They are localized field configurations
with finite energy which owe their existence not to topological properties,
but rather to the existence of a conserved charge. 
The first examples were studied by R. Friedberg, T. D.
Lee and A. Sirlin in \cite{Friedberg:1976me,Lee:1991ax} and some time
later Coleman studied what he termed \textit{Q-balls} \cite{Coleman:1985ki}.
These solitons, as we shall discuss below, are built out of scalar fields
with an unbroken $U(1)$ symmetry. As is well known, Derrick's theorem
precludes the existence of time-independent soliton configurations in $d\gtr 1+1$ if only scalar fields are involved. However,
non-topological solitons, both the Friedberg \textit{et al.} and Q-balls,
circumvent Derrick's theorem by an explicit time dependence and owe their
stability to the conserved charge associated to the unbroken $U(1)$. Despite
of this similarity, they differ in some other respects. For instance, in the
case of the Friedberg \textit{et al.} type, there's a second symmetry which is
spontaneously broken due to the presence of a Higgs-like scalar field. They represent, as we shall see, 
regions of space in which the Higgs vacuum expectation value (VEV) is suppressed due to the presence
of heavy particles. In this work we concentrate solely in this kind of
Higgs-based solitons and the role of quantum corrections as we believe they
might be relevant in applications to cosmology, although we will often
compare our results to those previously obtained for Q-balls.

Regarding Q-balls, it was first shown in \cite{Coleman:1985ki} that they exist in the limit
of large charge $Q$, and it was later shown in \cite{Kusenko:1997ad} that these
classical Q-balls actually exist all the way down to $Q=1$. These small
Q-balls may play an important role in cosmological applications \cite%
{Kusenko:1997ad} and therefore it is important to determine the validity of
the classical approximation. Therefore, the role of quantum corrections to these small
Q-balls was explored in \cite{Graham:2001hr} by using the methods developed
in \cite{Farhi:1998vx} and it was found that although they are small in comparison to the classical energy, they are comparable to the binding energy for small $Q$. It was
finally concluded that quantum corrections (for typical
values of coupling constants) render small Q-balls ($Q\lesssim 7$) unstable.

  In contrast to all these works, we do not discuss hypothetical objects which
  can only appear in various extensions beyond the Standard Model,
  but address the range of stability and existence of the Multi-Quanta Higgs Bags (MQHB) in the SM itself. 
  We have in mind the heaviest objects of the SM, top quarks $\bar t,t$ and $W,Z$ bosons, to
which we  refer generically as ``quanta". Of course, bags with many top quarks, or $top-bags$ as we will call them,
are quite different from scalar-quanta bags due both to their fermionic nature and their coupling to the Higgs field. 
However, these issues, and the potential role of these bags in the electroweak baryosynthesis problem, will be addressed elsewhere \cite{paperI, paperII}. In this work, which is mostly methodical in nature, we first address the issue of whether quantum corrections can or cannot destabilize these objects.



The interest in the issue originated from the question whether a
sufficiently heavy SM-type fermion should actually exist as a different
state, in which it depletes the Higgs VEV around itself and thus is
accompanied by its own ``bag" \cite{JS}.  Although classically this seemed
to be possible, it was shown by Farhi \textit{et al.} \cite{Farhi:2003iu} that
in fact quantum (one loop) effects destabilize such bags, except at such a
large coupling that the theory itself is apparently \textit{sick}, with an
instability of its ground state. The issue rest dormant for some time till
Nielsen and Froggatt \cite{Froggatt:2008ns} argued
that although the Yukawa coupling may still be small,
there should be binding when the number of particles $N$ is such that 
\begin{equation}
\frac{g^2}{4\pi} \ll 1, \,\,\, \frac{g^2}{4\pi} N > 1.
\end{equation}
Indeed, a similar generic argument explains for example why gravity, although very weak, 
can create bound states -- planets and stars -- but only for an ``astronomically large"   $N$. 

These authors also 
suggested that 12 top+antitop quarks (the ``magic number" corresponding to the
maximal occupancy of the lowest $S_{1/2}$ orbital, 6 quarks and 6
anti-quarks with all spin and color values) already form deeply bound states
with near-zero total mass. Here is when one of us (ES) became involved: in a
paper with Kuchiev and Flambaum \cite{Kuchiev:2008fd} we checked this
suggestion and showed that, unfortunately, it is $not$ the case.
While for a massless Higgs we found a \textit{weakly} bound state of 12
tops, for a realistic Higgs mass there is no bound state at all. 
A possible direction to consider is that of introducing new hypothetical fermions with a stronger coupling (larger masses) than that of the top quark and the possibility of these forming \textit{baryons}, well-bound by the Higgs. This issue has been discussed further in the second paper of the same authors \cite{Kuchiev:2008gt} and we will discuss the implications of our study for this later.

Although we disagreed with the calculation by Nielsen and Froggatt \cite{Froggatt:2008ns} on 12 tops,  we think that further studies of multi-quanta Higgs bags are well justified.  
In this paper, we will be especially interested in the case in which the number
of ``quanta" $N$ is large enough to considerably modify the Higgs VEV compared to its vacuum value. In other words, we will focus on relativistic
bound states, in which the ``quanta" inside the soliton have masses quite
different from their vacuum ones, rather than non-relativistic atom-like
bound states with smaller $N$. This sector of the SM spectroscopy constitutes
an interesting class of relativistic many-body systems, quite different from atoms and
nuclei in many respects, especially regarding quantum corrections, which we study.
While the experimental production of a large number of  heavy
quarks/bosons in a sufficiently small volume is clearly impossible,  at the LHC
or any other future proton accelerator, such states may play a significant role in
cosmological baryogenesis at temperatures slightly below the electroweak phase transition \cite{paperII}. 

Turning now to the main subject of this paper; the role of \emph{quantum corrections} to the
energy of such configurations, we express their energy generically as a quantum loop diagram
expansion  
\begin{equation}  \label{eq_Ck}
E(N)=E_{classical} \left[ 1+ \sum_k C_k(N) (\frac{g^2}{4\pi})^k \right].
\end{equation}
Below we calculate the one-loop coefficient $C_1$ closely following both
the specific model and the practical methods developed in \cite{Farhi:1998vx}.  Very specifically, the issue we investigated is the $N$-dependence of the
1-loop radiative correction. 

The paper is structured as follows. In Section \ref{sec_classical} we
formulate the Lagrangian and the \textquotedblleft thin wall"
approximation used to prove the existence of solitons at large enough $N$.
In Section \ref{sec_quantum} we review the methods developed in \cite%
{Farhi:1998vx} which provide a practical tool for numerically calculating
the 1-loop radiative corrections to soliton configurations. It is important
to note that these methods were designed for time-independent classical
configurations while the non-topological configurations we just discussed
have an explicit time dependence by having a non-zero $N$. However, as shown
in \cite{Graham:2001hr} (in an application to \textit{Q}-balls) the method
can easily be adapted to study a time-dependent configuration as well. After
an estimate of the effect of quantum corrections in Section \ref{quantum
estimate} and a discussion of other secondary calculational details in
Section \ref{sec_calc}, we proceed to our results in Section \ref%
{sec_results}.

\section{A classical soliton}

\label{sec_classical}

Consider the action 
\begin{equation}
S=\int d^{4}x\left( \frac{1}{2}(\partial _{\mu }\phi )^{2}-\frac{\lambda }{4!%
}(\phi ^{2}-v^{2})^{2}+\left\vert \partial _{\mu }\psi \right\vert
^{2}-g\psi ^{\ast }\phi ^{2}\psi \right)  \label{action}
\end{equation}%
where $\phi $ is a real scalar field, modeling the SM Higgs, with its VEV $%
\phi (r=\infty )=v$. A complex scalar field $\psi$ would have ``quanta", our
simple version of $t,W$ and $Z$. This system was studied in \cite%
{Friedberg:1976me} as an example which exhibits a soliton solution at the
classical level. In this section we review the general arguments which were
given in favor of the existence of these objects.

Due to the global symmetry $\psi \rightarrow e^{i\alpha }\psi $ in (\ref%
{action}), there's an associated conserved current, $\partial _{\mu }j^{\mu
}=0,$ with 
\begin{equation}
j_{\mu }=i\left( \psi ^{\ast }\partial _{\mu }\psi -\psi \partial _{\mu
}\psi ^{\ast }\right)
\end{equation}%
and a corresponding conserved particle number 
\begin{equation}
N=\int j^{0}d^{3}x.  \label{Number operator}
\end{equation}

Now, let the complex scalar be a time-dependent field rotating in internal
space with a frequency $\omega $, i.e. 
\begin{equation}
\psi (x,t)=e^{-i\omega t}\psi (x).
\end{equation}

The classical energy $E_{cl}$ from (\ref{action}) reads,

\begin{multline}
E_{cl}=\int d^{3}x\left[ \frac{1}{2}(\vec{\nabla}\phi )^{2}+\left\vert \vec{%
\nabla}\psi \right\vert ^{2}+\omega ^{2}\left\vert \psi \right\vert
^{2}\right.  \label{energy} \\
\left. +\frac{\lambda }{4!}(\phi ^{2}-v^{2})^{2}+g\phi ^{2}\left\vert \psi
\right\vert ^{2}\right]
\end{multline}

and from (\ref{Number operator}) we have 
\begin{equation}
N=2\omega \int d^{3}x\left\vert \psi \right\vert ^{2}.  \label{charge}
\end{equation}%
The equations of motion read 
\begin{eqnarray}
-\omega^{2}\psi -\vec{\nabla}^{2}\psi +g\phi ^{2}\psi &=&0,
\label{EOM for psi} \\
-\vec{\nabla}^{2}\phi -2g\left\vert \psi \right\vert ^{2}\phi +\frac{\lambda 
}{6}(\phi ^{2}-v^{2})\phi &=&0.
\end{eqnarray}%
By using the equation of motion (\ref{EOM for psi}) for $\psi$ in (\ref%
{energy}) and the expression (\ref{charge}) we find

\begin{equation}
E_{cl}=E_{h}+N\omega  \label{classical energy}
\end{equation}%
with

\begin{equation}
E_{h}=\int d^{3}x\left[ \frac{1}{2}\left\vert \vec{\nabla}h\right\vert ^{2}+%
\frac{m^{2}}{8v^{2}}(h^{2}+2vh)^{2}\right],  \label{Higgs Energy}
\end{equation}%
where we have defined as usual $h\equiv \phi -v$. Therefore, the classic
configuration consists of a static background field $h(x)$ and $N$ ``quanta"
of the $\psi $ field at an energy level $E_{0}=w$. In the Higgs vacuum, $%
\phi(r)=v$, $N$ quanta of the $\psi $ field would be forced to rest at the
bottom of the continuum spectrum and the total energy of the system would
simply be $NM $. However, in the background of a non-trivial Higgs field
there are two competing effects. On the one hand, the gradient and potential
terms in (\ref{Higgs Energy}) increase the energy but, on the other hand,
there might be some bound states levels with energy $0<E_{j}<M$ which can
allocate the quanta, lowering the energy of the system of particles at the
expense of creating such a Higgs configuration. In particular, if there's a
large region of nearly-zero Higgs VEV (where particles would be nearly
massless), a soliton will be possible if the energy increase by creating
such a region is offset by the energy decrease of $N$ particles giving away
their mass.

Indeed, the existence of such solitons was shown in \cite{Friedberg:1976me}
by considering the following trial configuration: A time-independent Higgs
field which vanishes inside a sphere of radius $R$ while outside it takes
its asymptotic value, with a transition region of size $l$. The complex
scalar $\psi (t,x)=e^{-i\omega t}\psi (x)$ is non-zero inside the sphere and
vanishes outside. In a \textit{thin-wall} approximation, $l\ll R,$ this
provides a spherical well in which the complex field forms a relativistic
bound state, i.e. 
\begin{equation}
\begin{array}{c}
\phi =0,\quad \psi (x,t)=e^{-i\omega t}\frac{A}{r}\sin (\omega r),\quad
r\lesssim R, \\ 
\phi =v(1-e^{-(r-R)/l}),\quad \psi =0,\quad r\gtrsim R.%
\end{array}
\label{solution}
\end{equation}%
with $\omega =\frac{\pi }{R}$. From (\ref{classical energy}), (\ref{Higgs
Energy}) the energy of such a configuration, in the limit of large $R$, reads%
\begin{equation}
E=\frac{\pi N}{R}+\frac{\pi m^{4}}{2\lambda }R^{3}+O(R^{2}),
\label{energy classical soliton}
\end{equation}%
which has a minimum at 
\begin{equation}
N\simeq \frac{3m^{4}R^{4}}{2\lambda },\quad E\simeq \frac{4\pi m}{3}\left( 
\frac{3}{2\lambda }\right) ^{1/4}N^{3/4},  \label{thin wall energy}
\end{equation}%
where $m^{2}=\lambda v^{2}/3$ and $M^{2}=gv^{2}$. Notice the power of $N$
with which the energy scales. If the Higgs field took its asymptotic value
everywhere $($i.e. $\phi (r)=v)$, the total energy of the system would
simply be given by $E=NM$. Since $E\propto N^{3/4}\ll N$ for large $N$,
there's always a large enough $N$ which makes the soliton configuration
energetically favored over simply having $N$ particles in the Higgs vacuum.
Notice that the critical configuration with $E_{c}=MN_{c}$ is given by%
\begin{equation}
N_{c}\simeq \left( \frac{4\pi m}{3M}\right) ^{4}\frac{3}{2\lambda }.
\label{critical bag}
\end{equation}%
Since $R$ grows with $N$, this expression becomes more accurate for large $N$%
, i.e. when $M/m\rightarrow 0$ expression (\ref{critical bag}) gives the
limiting value for $N_{c}$. It was similarly shown in \cite{Friedberg:1976me}
that in the limit $M/m\rightarrow \infty $, an upper bound on $N_c$ is given
by%
\begin{equation}
N_{c}< 336\left( \frac{m}{M}\right) ^{3}\frac{1}{\lambda }
\label{upper bound for Nc}
\end{equation}%
and therefore when the Higgs becomes massless the solitons exist not only for large enough $N$ but actually all
the way down to small $N$. 

Recall that Derrick's theorem precludes the existence of \textit{time-
independent} solitons with an action such as (\ref{action}), but this no-go
theorem is clearly circumvented here because of the explicit time dependence
in (\ref{solution}), which leads to $N\sim \omega A^{2}$. If we had no time
dependence the first term in (\ref{energy classical soliton}) would vanish,
rendering the system's minimum at $R=0$, in accordance with Derrick's
theorem.

Although we have shown the existence of classical bags in certain approximations, in order to explicitly find such configurations with minimal energy we adopted a variational strategy. We 
first fix a smooth, 1-parameter, bag-shaped, background Higgs field ansatz on
which we numerically solve the equation of motion for $\psi$. We then
minimize the resulting energy with respect to the variational parameter.
Once the classical solution is found we turn to the study of quantum
corrections by implementing the method developed in \cite{Farhi:1998vx}, which for completeness is
reviewed in the next section.

\section{Quantum Corrections: Generalities}

\label{sec_quantum}

\subsection{ Time-independent configurations}

As we have seen, a non-trivial Higgs background may create
classical bag-like solitons. However, it also has a secondary effect at the quantum
level, which is that of disrupting the energy levels of the vacuum by the
Cassimir effect. To study whether this affects or not the
existence of solitons we must study the vacuum energy, given to leading
order in $\hbar $, by the effective action. The calculation of such an
object will be plagued by the usual divergences of field theory and requires
proper renormalization. In this section we review the method developed by
Farhi \textit{et al.} \cite{Farhi:1998vx}, which precisely provides us a way of
calculating the effective action in a manner consistent with on-shell mass
and coupling renormalization. We shall consider the effective action induced
only by integrating out fluctuations in $\psi$.

The one-loop effective action, induced by integrating out the $\psi $ field,
is given by 
\begin{equation}
e^{iS_{eff}[\phi ]}=e^{iS_{\phi }}\frac{\int D\psi ^{\ast }D\psi e^{iS_{\psi
}}}{\int D\psi ^{\ast }D\psi e^{\left. iS_{\psi }\right\vert _{V=0}}}
\end{equation}%
and the effective energy is then  
\begin{equation}
E_{eff}=-\lim_{T\rightarrow \infty }\frac{1}{T}S_{eff},\quad
S_{eff}=S_{cl}+S_{ct}-\log \det H
\end{equation}%
where $H$ is the quadratic operator $\frac{\delta ^{2}S}{\delta \psi ^{\ast
}\delta \psi }$, generally of the form $H=-\square +M^{2}+V$. The matrix
identity $\log \det H=Tr\log H$ allows us to write the effective action as a
trace in functional space. We can perform this trace over a complete set of
functions satisfying $H\psi _{\alpha }=0$ or%
\begin{equation}
\left( -\nabla ^{2}+M^{2}+V\right) \psi _{\alpha }=E_{\alpha }\psi _{\alpha }
\label{eigenvalue equation with potential}
\end{equation}%
which in the free ($V=0$) case are simply plane waves $e^{ikx}$. Therefore,
in that case, $E_{vac.}^{\psi }\sim \int d^{4}k\log (k^{2}+M^{2})$ which by
standard methods can be brought into the form $\int d^{3}k\sqrt{\vec{k}%
^{2}+M^{2}}=\int d^{3}k\left\vert E_{k}\right\vert $, which is the usual
vacuum energy. In the presence of a background potential there might be,
in addition to the continuum, some discrete levels corresponding to
scattering bound states. In such case the vacuum energy generally reads

\begin{equation}
E_{vac.}^{_{\psi }}=\sumint_{\alpha }\left\vert E_{\alpha }\right\vert,
\end{equation}%
where $\alpha =\{i,k\}$ labels the discrete ($i$) eigenvalues, if any, and
continuous ($k$) eigenvalues and $E_{\alpha }$ are the energies of the
excitations. This is simply the sum of zero-point fluctuations $\frac{1}{2}%
\hbar w$ where the $\frac{1}{2}$ factor is absent due to $\psi $ being a
complex field. Therefore, there are 3 contributions to the total effective
energy given by%
\begin{equation}
E[h]=E_{cl}[h]+E_{ct}[h]+E_{vac.}^{_{\psi }}[h],
\end{equation}%
where $E_{ct}$ is the necessary renormalization energy counterterm. For a
fixed $h$, the spectrum consists of a continuous spectrum $E^{2}=k^{2}+M^{2}$
and (possibly) a finite number of bound states with energy $0<E_{j}<M$, i.e. 
\begin{equation}
E_{vac.}^{_{\psi }}[h]=\sumint_{\alpha }E_{\alpha
}=\sum_{j}E_{j}+\sum_{l}(2l+1)\int dk\rho _{l}(k)E(k)
\end{equation}%
where $\rho _{l}(k)$ is the density of states in momentum space and the $%
(2l+1)$ factor accounts for the degeneracy in the angular momentum
projection. It can be shown that \footnote{%
This is easily seen as follows: Consider a finite but large region of space
with radius $R$ and a potential $V$ with finite range. Asymptotically, the
eigenfunctions will be given by trigonometric functions with argument $%
kr+\delta_l(k)$. Imposing that the eigenfunctions vanish at the boundary we
find $kR+\delta_l(k)=n\pi$ and similarly for the free case, $V=0$. In the
continuum limit, $R\rightarrow \infty$, the states become infinitesimally
close and we have $dn=\frac{1}{\pi}(Rdk+d\delta_l(k))$. Letting $\rho=dn/dk$
we find the desired relation.} 
\begin{equation}
\rho _{l}(k)=\rho _{l}^{0}(k)+\frac{1}{\pi }\frac{d}{dk}\delta _{l}(k)
\end{equation}%
where $\delta _{l}(k)$ is the scattering phase shift of the $l$'th partial
wave under the potential $V(r)$ and $\rho _{l}^{0}(k)$ is
the density of states in the absence of potential. Therefore, in order to
calculate the whole 1-loop effective action we must calculate the phase
shifts $\delta _{l}(k)$. As is standard in quantum mechanics, a Born
approximation may be used. This represents an expansion in number of insertions of the potential $V(r)$ and in this model only diagrams with 1 and 2 insertions are divergent. Therefore, one defines the subtracted phase shift
\begin{equation}
\bar{\delta}_{l}(k)\equiv \delta _{l}(k)-\delta _{l}^{(1)}(k)-\delta
_{l}^{(2)}(k)
\label{subtractions}
\end{equation}%
where $\delta _{l}^{(1)}(k)$ and $\delta _{l}^{(2)}(k)$ are the first and
second Born approximations to $\delta _{l}(k)$. It is shown in \cite{Farhi:1998vx} that these two subtractions can be reintroduced into the action as divergent diagrams, which together with $E_{ct}[h]$ combine to
give a finite contribution denoted by $\Gamma _{2}[h]$. Therefore, we have%
\begin{multline}
E[h]=E_{cl}[h]+\Gamma _{2}[h] \\
+\frac{1}{\pi }\sum_{l}(2l+1)\int dk\left( \frac{d}{dk}\bar{\delta}%
_{l}(k)\right) E(k)+\sum_{j}\left\vert E_{j}\right\vert .
\label{a}
\end{multline}

A very useful theorem due to Levinson states that the number of
bound states $n_{l}$ with angular momentum $l$ in a given background
potential is given by%
\begin{equation}
n_{l}=\frac{1}{\pi }\left[ \delta _{l}(0)-\delta _{l}(\infty )\right] .
\label{Levinsons theorem}
\end{equation}%
This allows us to partially integrate $d/dk$ in (\ref{a}), picking up a boundary term $-M$
for each bound state, which finally gives an effective energy containing 4
terms, namely%
\begin{equation}
E[h]=E_{cl}[h]+E_{bound}[h]+E_{cont.}[h]+\Gamma _{2}[h]  \label{Energy}
\end{equation}%
where 
\begin{eqnarray}
E_{bound}[h] &=&\sum_{j}(\left\vert E_{j}\right\vert -M), \label{Bound States} \\
E_{cont.}[h] &=&-\frac{1}{\pi }\sum (2l+1)\int_{0}^{\infty }dk\bar{\delta}%
_{l}(k)\frac{k}{\sqrt{k^{2}+M^{2}}},
  \label{continuum energy} 
\end{eqnarray}%
and $E_{cl}$ is given by (\ref{classical energy}). Notice that after the
boundary term due to Levinson's theorem is included, the quantum correction
due to the bound states is negative and therefore it lowers the total energy. This raises the possibility that one might find quantum solitons
which classically are forbidden due to Derrick's
theorem. The existence of such solitons was explored
in \cite{Farhi:1998vx} with no success. From the previous discussions,
however, we know that classical, time-dependent, soliton configurations exist for $N>0$ and
we are interested in determining the effect of quantum corrections to these
classical configurations.
\newpage
\subsection{ Time-dependent configurations}
\label{Time-dependent configurations}
We saw above that the method developed in \cite{Farhi:1998vx} deals
specifically with time-independent classical configurations. Therefore, to
find the role of quantum corrections to the class of non-topological
solitons we are interested in, we must adjust the methods to a
time-dependent configuration. This has already been done in the case of $Q$%
-balls in \cite{Graham:2001hr} by one of the same authors who developed
the methods for the static case. In our case the same reasoning applies and
we closely follow the same derivation. Nevertheless, as we shall discuss, in applying this method to the case at hand there's an important difference with respect to Q-balls. 
The extension to the time-dependent case is easily carried out by studying fluctuations in a corotating frame. Let 
\begin{eqnarray}
\psi (x,t) &=&e^{-i\omega t}\left( \psi _{cl}(x)+\eta (x,t)\right) , \\
\phi (x,t) &=&\phi _{cl}(x),
\end{eqnarray}%
with $e^{-iwt}\psi _{cl}(x)$ the classical soliton discussed above and $%
\delta \psi =e^{-iwt}\eta (x,t)$ an arbitrary quantum fluctuation. Then the
classical action reads 
\begin{multline}
S_{cl}=\int d^{4}x\left[ -\frac{1}{2}(\vec{\nabla}\phi _{cl})^{2}-\left\vert 
\vec{\nabla}\psi _{cl}\right\vert ^{2}+\omega ^{2}\psi _{cl}^{2}\right.  \\
\left. -\frac{\lambda }{4!}(\phi _{cl}{}^{2}-v^{2})^{2}-g\left\vert \psi
_{cl}\right\vert ^{2}\phi _{cl}^{2}\right] 
\end{multline}%
and to second oder in quantum fluctuations of $\psi $ 
\begin{multline}
S_{\eta }^{(2)}=\int d^{4}x\left[ \left\vert (\partial _{0}-i\omega )\eta
(x,t)\right\vert ^{2}-\left\vert \vec{\nabla}\eta (x,t)\right\vert
^{2}\right.  \\
\left. -g\phi _{cl}^{2}\left\vert \eta (x,t)\right\vert ^{2}\right] .
\end{multline}%
Therefore, 
\begin{equation}
-(\partial _{0}-i\omega )^{2}+\vec{\nabla}^{2}-g\phi _{cl}^{2}
\end{equation}%
is the quadratic operator of quantum fluctuations. For convenience, we
parametrize $\eta (x,t)=e^{i(\omega -E)t}\eta (x)$, so that fluctuations
over the \textit{static} configuration $\left\{ \phi _{cl}(x),\psi
_{cl}(x)\right\} $ have an energy $E-\omega $ and hence their contribution
to the effective energy is 
\begin{equation}
E_{vac.}^{\psi }=\frac{1}{2}\sumint_{\alpha }\left\vert E_{\alpha }-\omega
\right\vert,
\end{equation}%
where the $E_{\alpha }$'s are the solutions to the eigenvalue problem 
\begin{equation}
\left( -E_{\alpha }^{2}-\vec{\nabla}^{2}+g\phi _{cl}^{2}\right) \eta
_{\alpha }(x)=0,  \label{eigenvalue problem}
\end{equation}%
i.e. we must study the zero modes of the operator $H=-E_{\alpha }^{2}-\vec{%
\nabla}^{2}+g\phi _{cl}^{2}$. Notice that we are left with a reduced problem
which is identical to that we would of have to solve for the static case (%
\ref{eigenvalue equation with potential}), with $V=g(h^{2}+2hv)\equiv g\chi $
and $M^{2}=gv^{2}$, but the contribution of each level is shifted by $\omega 
$. Now, since the spectrum in (\ref{eigenvalue problem}) is symmetric in $%
E\rightarrow -E$, we can write%
\begin{equation}
E_{vac.}^{_{\psi }}[h]=\frac{1}{2}\sum_{E_{\alpha }>0}\left\vert E_{\alpha
}-\omega \right\vert +\left\vert E_{\alpha }+\omega \right\vert
=\sum_{E_{\alpha }>0}\max \{E_{\alpha },\omega \}.
\end{equation}%
As we shall see below, \ if we only integrate out the fluctuations $\delta
\psi $ from (\ref{action}) we have $E_{\alpha }\geq \omega $ and therefore
the vacuum energy is again simply 
\begin{equation}
E_{vac.}^{_{\psi }}[h]=\sumint_{\alpha }\left\vert E_{\alpha }\right\vert 
\label{quantum energy 2}
\end{equation}%
and the total energy reads

\begin{equation}
E=E_{cl}+\sum_{i}\left( \left\vert E_{i}\right\vert -M\right) +E_{cont.}+
\Gamma_{2},  \label{variational energy}
\end{equation}%
with $E_{cl}$ given by (\ref{classical energy}). To
see that we have $E_{\alpha }\geq \omega $ in this case, we note that from
the equations of motion we find $\left( -\omega ^{2}-\vec{\nabla}^{2}+g\phi
_{cl}^{2}\right) \psi _{cl}(x)=0$ in the $l=$ $0$ channel, which is the most
tightly bound state and therefore the lowest positive energy is $%
E_{0}=\left\vert \omega \right\vert $, proving our assertion. It is important to note however that
had we also considered fluctuations of the Higgs field as well, there would
be a zero mode of the quadratic operator $H$ (which would be $2\times 2$)
now in the $l=1$ channel, corresponding to translations of the
soliton. This would imply that there's an energy level $E_{0}$ below $\omega $
(in the $l=0$ channel), giving an additional contribution to the effective
energy, namely $\omega -E_{0}$, with respect to ($\ref{quantum energy 2}$).
This is important in the case of the $Q$-ball where one has no other choice
than considering such a zero mode as there's only one field involved and
actually this term dominates the whole quantum correction \cite%
{Graham:2001hr}. This difference is important since this additional term, being positive, works\textit{\ against} the stability of
these solitons and is ultimately the responsible for rendering Q-balls with $Q\leq7$,  unstable. However, as just discussed, that's not the case here due
to the absence of the translational zero mode in the path integration.

\section{Estimate of quantum corrections}

\label{quantum estimate}

Before going to the numerical calculations, we can illustrate the effects of
quantum corrections on the solitons described in Section \ref{sec_classical}
by using the thin-wall construction. This configuration corresponds to a
spherical potential well of depth $-gv^{2}=-M^2$ and size $R$, i.e. $%
V=-M^2\Theta(R-r)$, and the evaluation of the quantum corrections requires
solving the energy levels of such a system. We wish to find the dependence
of these energy levels on the size of the soliton. We can easily do this for the continuum energy relative to the Higgs vacuum. It is
given by 
\begin{equation}
E_{cont.}=\int d^{3}x\int d^{3}k\left[ \sqrt{k^{2}+M^{2}+V}-\sqrt{k^{2}+M^{2}%
}\right]
\end{equation}%
and therefore 
\begin{equation*}
E_{cont.}=(4\pi )^{2}\int_{0}^{R}r^{2}dr\int dkk^{2}\left[ k-\sqrt{%
k^{2}+M^{2}}\right] .
\end{equation*}%
The integral over $k$ is obviously divergent and requires proper
regularization. However, the $R$-dependence is given by 
\begin{equation}
E_{cont.}\sim R^{3}\sim N^{3/4}
\end{equation}%
and therefore it grows as the volume of the soliton and its effect is that
of modifying the coefficient of $N^{3/4}$ in the classical energy (\ref{thin
wall energy}). The precise dependence of the bound states' contribution is
harder to find analytically and will depend strongly on the coupling $g$, as
this determines the depth of the potential $-gv^{2}$. Nevertheless, we
can understand its main effect. We know that a spherical well of depth 
$-gv^2$ has a minimum size $R_0=\frac{\pi}{2v\sqrt{g}}$ at which the first bound
state appears. For large coupling our potential becomes deeper and a
large number of states could is bound. As is well known, in the case of an
infinite spherical well there are an infinite number of bound states given
by the spherical Bessel functions $j_{l}(kr)$. The boundary condition $%
j_{l}(kR)=0$ determines therefore the energy levels of a relativistic particle to be 
\begin{equation}
E_{nl}=\frac{1}{R}\beta _{nl},
\end{equation}%
where $\beta _{nl}$ represents the $n$'th zero of the $l$'th order Bessel
function. The number $n(R)$ of bound states in a \textit{finite} potential
well is finite and grows with $R$ and therefore the contribution (%
\ref{Bound States}) will be given by $n(R)(1/R-M)$, which may be significant
at large $R$. Indeed, as we shall comment more later, for large enough
coupling, it may become so negative that it wins over the classical energy
of the soliton, making the vacuum unstable under the creation of empty bags
of zero Higgs VEV. For relatively small values of $g$ ($\sim 1-3$) this term
has the finite effect of lowering the number $N_c$ of particles required to
have a stable configuration.

\section{ Calculational Details}

\label{sec_calc}


\subsection{ Variational method for classical solution:}

Quantum corrections require some operations with the background Higgs field
(such as finding its Fourier transform) which is difficult in practice to do
for numerically generated shapes. Therefore we adopted a variational
approach, in which the field is given in a simple analytic form, with
certain parameters to be determined from energy minimization.

Our purpose is to find a variational minimum of (\ref{variational energy}). For this we used a family of 1-parameter Gaussian backgrounds for the Higgs%
\begin{equation}
h(r)=-ve^{-r^{2}v^{2}/2w^{2}}
\end{equation}%
where $w>0$ is the variational parameter representing the size of the configuration. Notice that $\left\langle \phi \right\rangle \sim 0$ for $r\lesssim w$ while $\left\langle \phi \right\rangle \sim v$ for $r\gtrsim w$%
. As is customary, we refer to this configuration as a \textit{bag}. It
represents a region of size $\sim w$ of symmetric phase embedded in a space
of broken phase. Inserting this ansatz into (\ref{Higgs Energy}) gives us an
analytical expression for the Higgs energy, which we denote $E_{bag}(w)$.
Notice also that this profile for the Higgs provides an attractive potential $g \chi
(r)$ under which the field $\psi $ scatters. By solving equation (\ref%
{eigenvalue problem}) with $l=0$ we find $\omega$, which is the lowest bound
state level, henceforth denoted by $E_{0}$.

Recall that the energy of the soliton is given by 
\begin{equation}
E(w)=E_{bag}(w)+NE_{0}(w).
\end{equation}%
For the energy to have a minimum we require that $(\delta E)_{N}=0$ and
therefore%
\begin{equation}
E_{bag}^{\prime }(w)+NE_{0}^{\prime }(w)=0
\end{equation}%
with $^{\prime }$ denoting derivative w.r.t. $w$. This defines $N$ as a
function of the bag's size $w$ by 
\begin{equation}
N(w)=-\frac{E_{bag}^{\prime }(w)}{E_{0}^{\prime }(w)}.
\label{expression for N}
\end{equation}%
Furthermore, we must also have $E_{bag}^{^{\prime \prime
}}(w)+NE_{0}^{^{\prime \prime }}(w)>0$. By taking the derivative of the
first equation and noting that $E_{0}^{^{\prime }}<0$, we find%
\begin{equation}
N^{\prime }(w)>0  \label{condition on N}
\end{equation}%
to have a minimum. We are
therefore interested to see if there's a range for which $N^{\prime }(w)>0$
with $N$ given by (\ref{expression for N}).

\subsection{Quantum Corrections}

\bigskip
 As explained in the Introduction we will follow here \cite{Farhi:1998vx}, who developed a rather practical way to calculate the quantum correction, due to one-loop diagrams involving the ``matter field" $\psi$, to the total energy of the system. It is given by several terms; the binding energy of the bound states, $%
E_{bound}$, the continuum term, $E_{cont.}$, representing regularized scattering phases, and the remaining regularized mass and charge renormalization, $%
\Gamma _{2}$. 

To evaluate $E_{bound}$ we proceed as follows: Recall that $\delta
_{l}(\infty )=0$ and therefore, according to Levinson's theorem (\ref%
{Levinsons theorem}), the number of bound states with angular momentum $l$
is given by%
\begin{equation}
n_{l}=\frac{1}{\pi }\delta _{l}(k=0).
\end{equation}%
%
According to \cite{Farhi:1998vx}, from the radial equation of motion, one finds that the phase shifts are given by $\delta
_{l}(k)=-2\Re \beta _{l}(k,r=0)$ with the $\beta _{l}$'s \ satisfying 
\begin{equation}
-i\beta _{l}^{^{\prime \prime }}-2ikp_{l}(kr)+2(\beta _{l}^{^{\prime }})^{2}+%
\frac{1}{2}g\chi (r)=0  \label{part wave diff eq}
\end{equation}%
where $p_{l}(x)\equiv d/dx\left( \ln \left[ xh_{l}^{1}(x)\right] \right) $
and $h_{l}^{1}(x)$ is a spherical Hankel function and $^{\prime }$ denotes
derivative with respect to $r$. By solving this differential equation and
taking the limit $k\rightarrow 0$ we found the number of bound states to be
expected in every angular momentum channel at a given $w$. With this
knowledge, we used a shooting method to find the energies of such states by
solving the differential equation%
\begin{equation}
\left( -\vec{\nabla}^{2}+g\phi _{cl}^{2}\right) \psi =E^{2}\psi .
\end{equation}%
That is, we repeatedly solved this differential equation for $\psi=\frac{u}{r%
}$ in a box of size $L=30$ for increasing values of $E$ from $0$ to $M$,
with boundary conditions $u(0)=0$, $u^{\prime}(0)=1$. We then selected those
solutions which vanish at infinity (i.e. $u(L)=0$), verifying that the
number of such solutions matched those predicted by Levinson's
theorem. 

To evaluate the continuum contribution (\ref{continuum energy}), one must
find the Born approximation to the phase shifts. These are similarly found
by solving a set of coupled differential equations, which are detailed in
the Appendix. 

The calculation of the renormalization term $\Gamma _{2}$ is detailed in the Appendix and
involves evaluating a double integral which is readily done numerically. It will be shown however that it is numerically suppressed to be several orders of magnitude smaller than the other contributions and may be neglected.
 
\bigskip

\section{Results}

\label{sec_results}
\begin{figure}[tbp]
\centering
\includegraphics[width=3.0in]{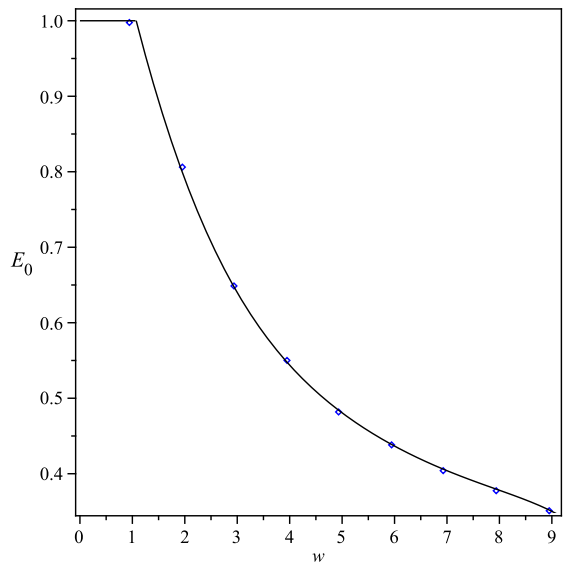}
\caption{Energy of the lowest bound state as a function of $w$ for $g=%
\protect\lambda=1$. The points show the results of our numerical
calculations while the continuous line is an interpolation which allows us
to find a smooth expression for $N$. We can see that there's a minimal value 
$w_{min}\sim 1$ at which a bound state appears and its energy decreases smoothly
to zero as $w$ increases. The number of levels in each channel coincides
with those predicted by Levinson's theorem.}
\label{ELowd=1w10}
\end{figure}

We discuss the results of our
  numerical calculations starting with the lowest s-wave bound state. Its dependence on the bag size $w$ 
 is shown in Fig. \ref{ELowd=1w10}. The points represent the numerical results and the continuous line is an
interpolation which, from (\ref{expression for N}), allowed us to find a
smooth expression  to be used below. Note that the curve does not go through the $w=0.9$ point, in which case there is no bound state: the point simply shows the  mass of the quantum $outside$ the bag, $E_0=M=1$
for the coupling we use.
Obviously, there's a minimum bag size $w_{min}\sim1$ at which the first state gets detached from the continuum to become a bound state with an energy just slightly below $M$. It is also clear that such shallow levels should be described by an atomic-like approach rather than by collective bags: we will not investigate their properties in this work, focusing only on deeper bound ones.
 
\begin{figure}[t]
\centering
\includegraphics[width=3.0in]{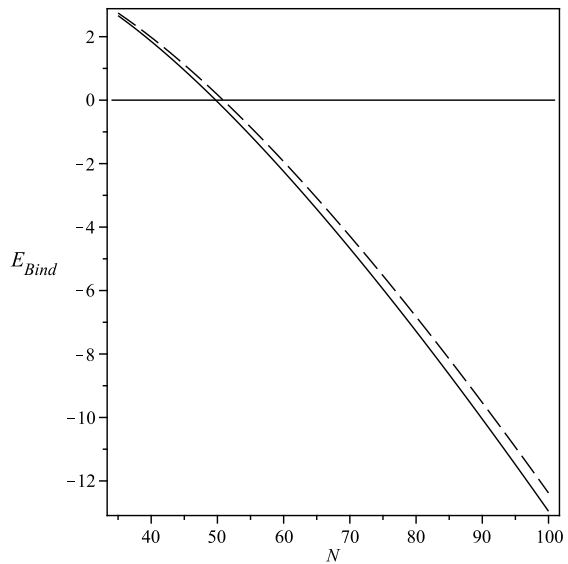}
\caption{This figure shows the binding energy of the soliton with $g=1$ as a
function of the number of particles. The solid line represents the binding
energy including the quantum correction, while the dashed line contains only
the classical energy. We can see that after a critical point $N_{c} \simeq52$,
it is energetically favorable to bind particles in a finite size bag rather
than contracting the bag to zero size forcing the particles to migrate to
the lowest state of the continuum spectrum with energy $NM$.}
\label{EnergyFunctionNw10d=1}
\end{figure}

Table \ref{table} shows some results for the energy of the bound
states in different $l$-channels for different sizes of the bag. As we increase $w$, new levels appear and the energy of the lowest bound state approaches $0$. We also observe that the levels appear in the same order as in the spherical well potential (see, e.g. \cite{Landau}). For instance, for $w=7.9$ the order of the energy levels is $1s,$ $1p,$ $1d,$ $2s,$ $2p,$ $2d,$ $3s$.
\begin{table}[tbp]
\begin{tabular}{|c||c|c|c|c|c|c|c|c|}
\hline
$l/w$ & $0.9$ & $1.9$ & $2.9$ & $3.9$ & $4.9$ & $5.9$ & $6.9$ & $7.9$ \\ 
\hline
$0$ & $-$ & $.809$ & $\left. 
\begin{array}{c}
.652 \\ 
.997%
\end{array}%
\right. $ & $\left. 
\begin{array}{c}
.552 \\ 
.892%
\end{array}%
\right. $ & $\left. 
\begin{array}{c}
.485 \\ 
.803%
\end{array}%
\right. $ & $\left. 
\begin{array}{c}
.442 \\ 
.745%
\end{array}%
\right. $ & $\left. 
\begin{array}{c}
.406 \\ 
.710%
\end{array}%
\right. $ & $\left. 
\begin{array}{c}
.380 \\ 
.682 \\ 
.982%
\end{array}%
\right. $ \\ \hline
$1$ & $-$ & $-$ & $.857$ & $.738$ & $\left. 
\begin{array}{c}
.652 \\ 
.935%
\end{array}%
\right. $ & $\left. 
\begin{array}{c}
.596 \\ 
.880%
\end{array}%
\right. $ & $\left. 
\begin{array}{c}
.552 \\ 
.846%
\end{array}%
\right. $ & $\left. 
\begin{array}{c}
.515 \\ 
.822%
\end{array}%
\right. $ \\ \hline
$2$ & $-$ & $-$ & $-$ & $.886$ & $.791$ & $.725$ & $\left. 
\begin{array}{c}
.682 \\ 
.977%
\end{array}%
\right. $ & $\left. 
\begin{array}{c}
.644 \\ 
.951%
\end{array}%
\right. $ \\ \hline
$3$ & $-$ & $-$ & $-$ & $-$ & $-$ & $-$ & $-$ & $-$ \\ \hline
\end{tabular}%
\caption{This table shows a typical set of bound state levels in different
angular momentum channels for different sizes of the bag with $g=\protect%
\lambda =1$.}
\label{table}
\end{table}
\begin{figure}[h]
\centering
\includegraphics[width=3.0in]{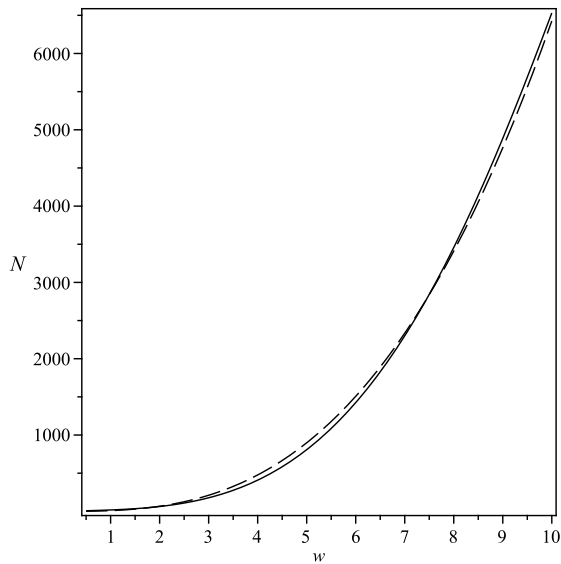}
\caption{Number of particles necessary to extremize the energy of a
configuration of bag's size $w$. The number of particles to create a minimal
bag is $\sim 20$, with a size $w_{min}\sim1$. The dashed line represents the
asymptotic behavior $\sim w^{2.86}$ for large $w$.}
\label{Nw}
\end{figure}
The number of quanta to bind in order to ensure that a bag of width $w$ has a saddle point is shown in Fig. \ref{Nw}. It  continuously increases with $w$: as we showed in the previous section the condition $N^{\prime }(w)>0$ needs to be satisfied in order
for the saddle point to be a local minimum and as we can see from the
figure, this is always the case for $w>w_{min}$. 
By taking $g=\lambda=1$ in equation (\ref{critical bag}), we expect a stable soliton to be found for $N\simeq51$.
Indeed, our results in Fig. \ref{EnergyFunctionNw10d=1} show that
without taking into account quantum corrections $N_{c} \simeq 52$ and the
effect of quantum corrections is to lower $N_c$ to $ \simeq50$. Recall that
we are ignoring the effect of the Higgs self-coupling at the quantum level
and therefore $\lambda$ enters only in the expression for the classical
energy. By taking $m/M\rightarrow 0$ (i.e. a BPS-like limit), the Higgs becomes
massless and therefore the exchange forces between quanta become Coulombic.
In this limit the minimum number of quanta $N_c$ to form a bound state is
expected to be significantly decreased, as confirmed by equation (\ref{upper
bound for Nc}). Notice also that this limit can be taken while keeping $g$,
and therefore quantum corrections, small. 
Our numerical results for $%
\lambda=10^{-4}$ and $g=3$ are shown in Fig. \ref%
{BindingEnergyLambda=0}. 
 By further increasing the coupling $g$ while
keeping $\lambda  \simeq 0$ the size of the critical
bag can be significantly decreased. For what we have seen, quantum corrections would not seem to spoil the
existence of such small solitons but rather promote it, contrary to the case of \emph{Q-balls}, due to the absence of the translational zero mode as discussed in \S \ref{Time-dependent configurations}.
\begin{figure}[h]
\centering
\includegraphics[width=3.0in]{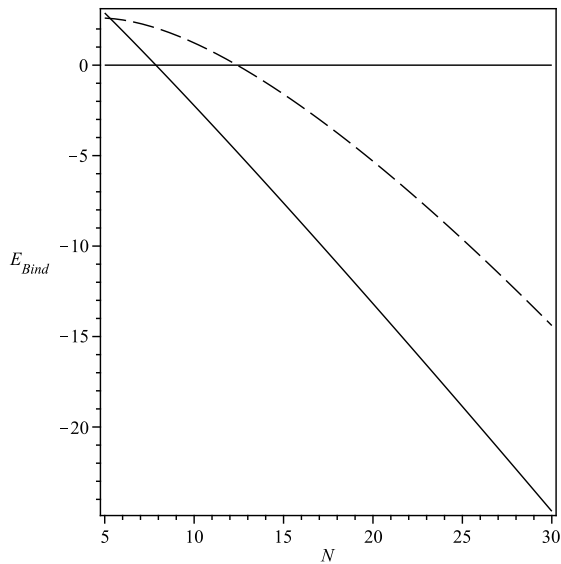}
\caption{This figure shows the binding energy of the soliton with $\protect%
\lambda=10^{-4}$ and $g=3$ as a function of the number of particles. Again,
the dashed line is the classical binding energy while the solid line
includes the quantum corrections. The number of particles to stabilize the
bag is decreased by a smaller Higgs self-coupling and quantum corrections
still favor the existence of such a soliton.}
\label{BindingEnergyLambda=0}
\end{figure}

\begin{figure}[t]
\centering
\includegraphics[width=3.0in]{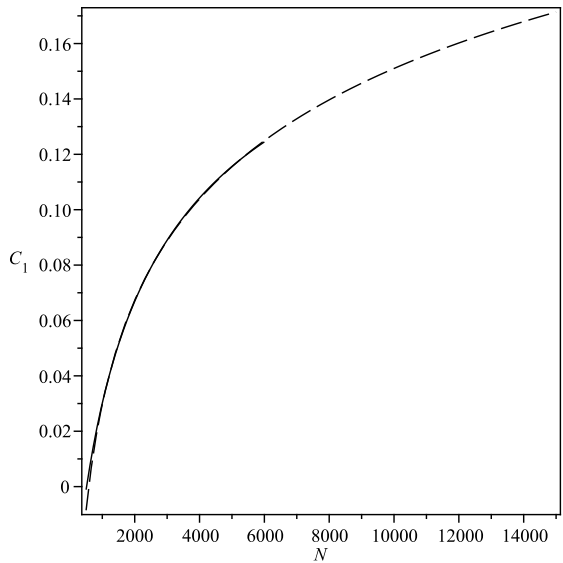}  
\caption{Coefficient $C_1(N)$ for $g=\protect\lambda=1$. The dashed line
represents an extrapolation of our results (solid line).}
\label{fig:C1g=1}
\end{figure}

The vacuum energy, relative to the classical energy is embodied in
the coefficient $C_1(N)$, which is shown in Fig. \ref{fig:C1g=1}.
Based on our discussion in Section \ref{quantum estimate}, we expect $C_1(N)=%
\frac{4\pi}{g^2}\frac{E_{vac.}}{E_{cl.}}$ to behave as $a-\frac{b}{f(N)}$ for large $N$,
with $a$ determined by the (renormalized) contribution from the continuum, while $b$ and $f(N)$ are determined by the bound-states contribution. Indeed, we found that our
results up to $N\sim 6000$ are nicely fit by $a=1.64$, $b={2.04}$ and a very small power ${%
f(N)=N^{0.03}}$: A logarithmic fit is also possible. Since this plot is for $g^2/4\pi\approx .08$, we conclude that the
one-loop coefficient $C_1$ in the generic expansion (\ref{eq_Ck}) grows
slowly with $N$ and is asymptotically of order $1$.
Therefore, even for the rather large coupling of top quarks ($g_t\approx 1$), quantum corrections seem to be under good theoretical control.

\begin{figure}[h]
\centering
\includegraphics[width=3.2in]{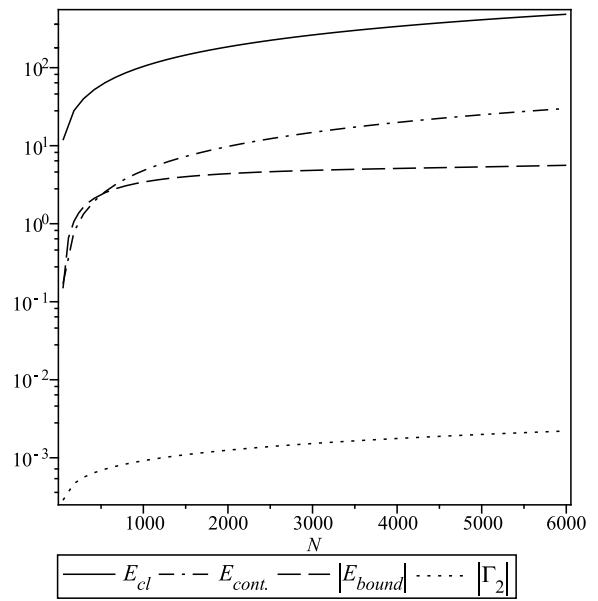}
\caption{All $4$ contributions to the total energy in a
log scale. $E_{bound}$ and $\Gamma_2$ are \textit{negative} so their
absolute values are shown. We can see that the main quantum correction for
small solitons is given by $E_{bound}$ while for large solitons $E_{cont.}$
becomes the most relevant. The renormalization term $\Gamma_2$ is
numerically suppressed by several orders of magnitude.}
\label{AllEnergyTerms1}
\end{figure}

\section{Summary and Discussion}
The  main content of the present paper is the explicit calculation of the regularized one-loop quantum correction to the 
energy of Multi-Quanta Higgs Bags, in a model in which ``quanta" are represented by a charged scalar field. The main results are:\\ (i) For the range of couplings corresponding to the heaviest fermion (the top quark)
and gauge bosons of the SM such bound states exist for $N$ larger than about 50; 
\\ (ii) quantum corrections
to their energy is reasonably small, at the level of few percents at large $N$, although 
they play a larger role at smaller $N$, where they tend to $lower$ the binding boundary $N_c$.
We do not think that the region close to it is under firm theoretical control, unlike the domain of large $N>100$. 
\\ (iii) For large solitons, the main
quantum correction is given by the scattering states continuum, which grows with $N$ as the classical energy itself. 
\\ (iv)  Even extrapolated to very large $N$ we find that the limits of validity of the
perturbative expansion do include  $g\approx 1$, i.e. the top quark of the SM. 

\begin{figure}[h]
\centering
\includegraphics[width=3.2in]{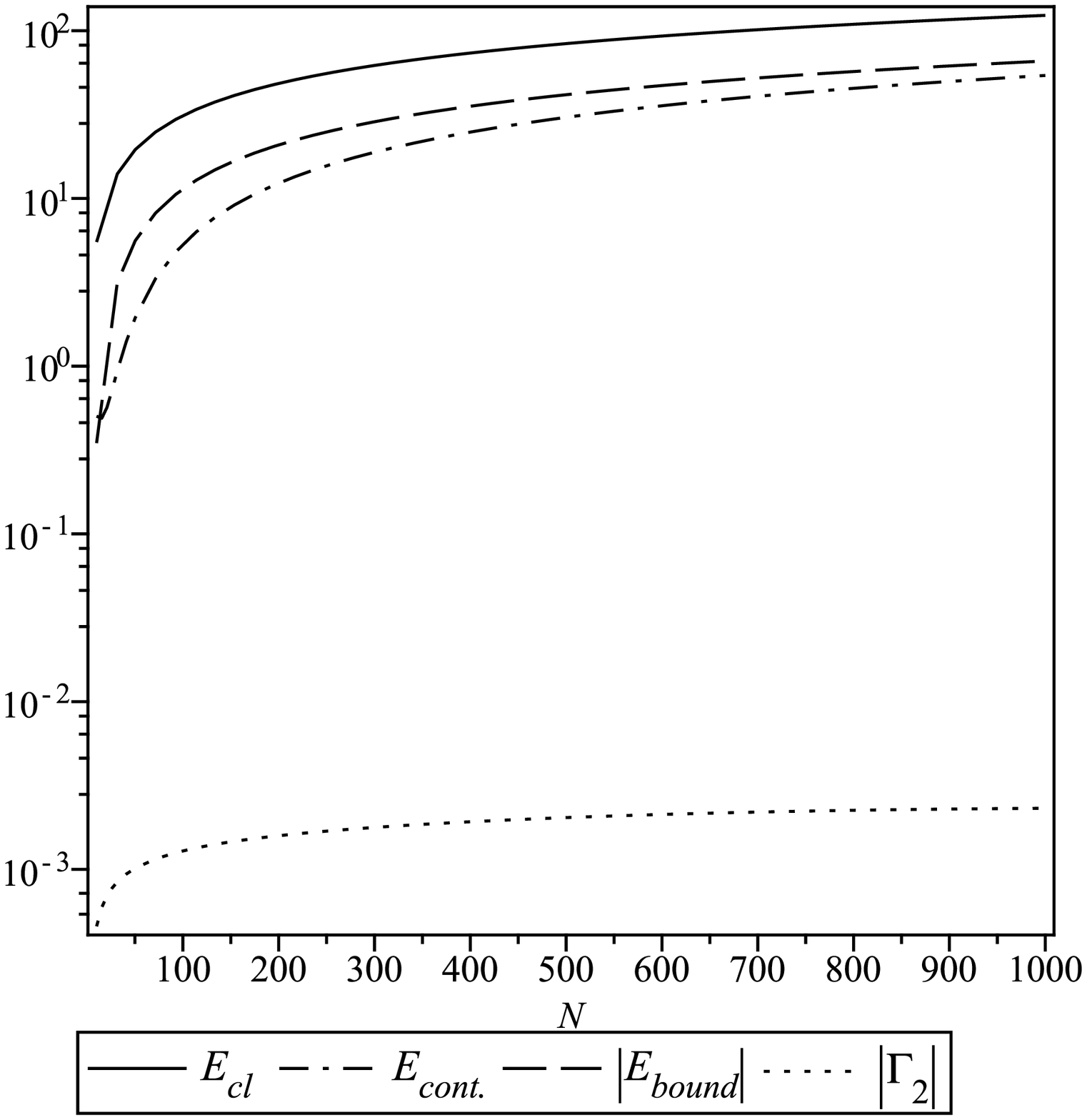}
\caption{All $4$ contributions to the total energy for $g=3$. For larger
coupling we can see how the contribution from $E_{bound}$ increases roughly
an order of magnitude. It becomes comparable in magnitude to $E_{cont.}$
(although with the opposite sign) for large $N$ and dominant at small $N$.}
\label{AllEnergyTerms_g=3}
\end{figure}

As mentioned in the Introduction, the modification of the (quantum bound states and classical bag) calculation for fermions (real top quarks) and
gauge bosons W,Z of the SM will be carried out elsewhere \cite{paperI}, together with cosmological applications \cite{paperII}.


As a remaining discussion issue, we may end up with the question when and what happens if the coupling is $larger$ than that of the top quark, namely $g > 1$.  
It has been argued by Farhi \textit{et al}. \cite{Farhi:1998vx} that for sufficiently large $g$ the model
becomes \textit{sick}, because the vacuum
itself (i.e. the $N=0$ state) becomes unstable against the spontaneous
creation of large empty bags. Our experience with the large-coupling domain confirms it. A closer look at each contribution to the
vacuum energy for larger $g$ uncovers the reason for this 
instability. From Figs. \ref{AllEnergyTerms1} and \ref{AllEnergyTerms_g=3} we can discern the contribution to the total energy from each separate term
in the effective energy for $g=1$ and $g=3$, respectively. For $g=1$ we observe that $\Gamma _{2}$ is numerically suppressed for all $N$ and the main quantum correction is due to an interplay between $E_{bound}$ and $E_{cont.}$, the former being dominant (and negative) for small $N$ ($\lesssim 100$),
while the latter becomes dominant (and positive) for large $N$ and behaves
similarly to the classical energy, as expected from our discussion in
Section \ref{quantum estimate}. Already for $g=3$ we found a significant increase in the contribution  $E_{bound}$,
which  becomes comparable in magnitude (but opposite in sign) to the classical
energy. Note that such coupling would correspond in the SM to fermions with a mass $\sim .5 \,TeV$,
and e.g. baryons made of such hypothetical fermions were suggested in \cite{Kuchiev:2008gt}.  We conclude
that those states, if such fermions do exist, would have noticeable quantum corrections. Thus their theoretical status is much less solid than that of top-balls, perhaps asking for dedicated further study.

For  $ g\gtrsim 5$  the bound energy term under discussion becomes so negative that it
completely compensates the classical energy of even an empty bag, therefore rendering the
vacuum unstable. However, at such strong coupling the theory is in serious trouble and clearly beyond the range of validity of any perturbative  expansion.



\vskip .25cm \textbf{Acknowledgments.} \vskip .2cm One author (ES)
acknowledges an inspiring discussion with Prof. H. Nielsen, which triggered
this project, as well as collaboration on a part of it with Profs. Flambaum and Kuchiev. This
work was supported in part by the US-DOE grant DE-FG-88ER40388.

\begin{figure}[h]
\centering
\includegraphics[width=3.0in]{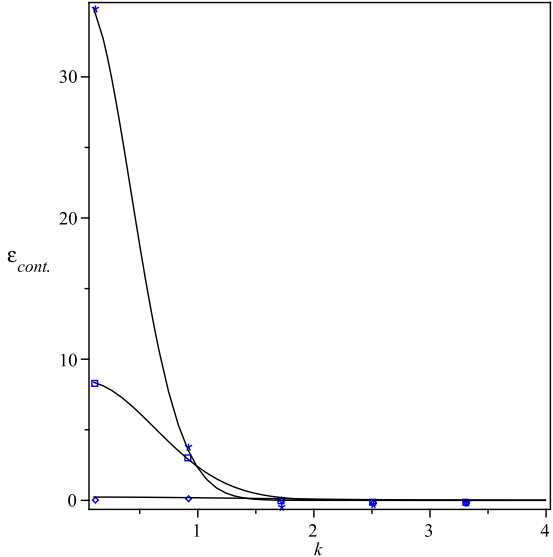}
\caption{Typical $k$-dependence of the Born-subtracted continuum energy for $%
3$ values of increasing $w$ from the bottom up. The integral over $k$ grows
quickly with increasing $w$.}
\label{PhaseShiftsEnergy}
\end{figure}

\bigskip
\section*{Appendix}


\bigskip Here we summarize the equations needed for the
calculation of the Born approximation to the phase shifts. This is needed
for the explicit calculation of the continuum contribution. We denote the
first and second Born terms by $\delta_{l1}$ and $\delta_{l2}$. A series
expansion in powers of $g$ for the phase shifts may be written as 
\begin{equation}
\beta _{l}=g\beta_{l1}+g^{2}\beta _{l2}+...
\end{equation}%
with $\delta_{l1}=-2g Re \beta_{l1}(k,r=0)$ and $\delta_{l2}=-2g^{2}
Re \beta_{l2}(k,r=0)$ and $\beta _{l1,2}$ satisfy the set of
coupled differential equations 
\begin{subequations}
\label{part wave Born diff eq}
\begin{eqnarray}
-i\beta _{l1}^{^{\prime \prime }}-2ikp_{l}(kr)\beta _{l1}^{^{\prime }}+\frac{%
1}{2}\chi (r) &=&0, \\
-i\beta _{l2}^{^{\prime \prime }}-2ikp_{l}(kr)\beta _{l2}^{^{\prime
}}+2(\beta _{l1}^{^{\prime }})^{2} &=&0.
\end{eqnarray}%
The continuum contribution is given by 
\end{subequations}
\begin{equation}
E_{cont.}=\int_{0}^{\infty }\varepsilon_{cont.}(k)dk,
\end{equation}%
with 
\begin{equation}
\varepsilon_{cont.}(k)=-\frac{1}{\pi }\sum_l (2l+1)\bar{\delta}_{l}(k)\frac{k%
}{\sqrt{k^{2}+M^{2}}}
\end{equation}%
and $\bar{\delta}_{l}(k)\equiv \delta _{l}(k)-\delta _{l}^{(1)}(k)-\delta
_{l}^{(2)}(k)$. Each of these terms is given as a solution to the set of
differential equations (\ref{part wave diff eq}) and (\ref{part wave Born
diff eq}). In figure (\ref{PhaseShiftsEnergy}) we show the continuum
energy $\varepsilon_{cont.}$ as a function of $k$ for 3 increasing values of $%
w$ from the bottom up. It can be seen that the Born subtracted $\delta _{l}$
is rapidly convergent in $k$ and that the integral grows quickly with $w$.

The renormalization effect, encoded in $\Gamma _{2}$ is calculated from a
set of Feynman diagrams and is given by

\begin{multline}
\Gamma _{2}=\frac{g^{2}}{(4\pi )^{2}}\int \frac{q^{2}}{2\pi ^{2}}%
dq\int_{0}^{1}dx\left\{ -\frac{4x(1-x)v^{2}q^{2}}{M^{2}-m^{2}x(1-x)}%
h(q)^{2}\right.  \notag \\
\left. +\left[ \log \frac{M^{2}+q^{2}x(1-x)}{M^{2}-m^{2}x(1-x)}-\frac{%
m^{2}x(1-x)}{M^{2}-m^{2}x(1-x)}\right] \chi (q)^{2}\right\}
\end{multline}
where $\chi (q)$ and $h(q)$ are the spatial Fourier transforms of $\chi (x)$
and $h(x)$. Notice that there are numerical factors $(4\pi )^{4}$ in the
denominator, which suppresses this contribution with respect to the other
ones. See \cite{Farhi:1998vx} for more details.


\end{document}